\documentclass{elsarticle}

\usepackage{color}
\usepackage{comment}
\usepackage[dvipdfmx]{hyperref}
\usepackage{lineno,hyperref}
\usepackage{lscape}
\journal{Journal of \LaTeX\ Templates}

\biboptions{authoryear}

\begin{document}

\begin{frontmatter}

\title{Impact degassing and atmospheric erosion on Venus, Earth, and Mars during the late accretion}

\author[mymainaddress]{Haruka Sakuraba\corref{mycorrespondingauthor}}
\cortext[mycorrespondingauthor]{Corresponding author}
\ead{sakuraba@eps.sci.titech.ac.jp}

\author[mysecondaryaddress]{Hiroyuki Kurokawa}
\author[mysecondaryaddress]{Hidenori Genda}

\address[mymainaddress]{Department of Earth and Planetary Sciences, Tokyo Institute of Technology, Ookayama, Meguro-ku, Tokyo, 152-8551, Japan}
\address[mysecondaryaddress]{Earth-Life Science Institute, Tokyo Institute of Technology, Ookayama, Meguro-ku, Tokyo, 152-8550, Japan}

\begin{abstract}

The atmospheres of the terrestrial planets are known to have been modified as a consequence of the impact degassing and atmospheric erosion during the late accretion. 
Despite the commonality of these processes, there are distinct gaps -- roughly two orders of magnitude -- between the abundances of noble gases and nitrogen in the present-day atmospheres on Venus, Earth, and Mars. The element partitioning on planetary surfaces is thought to be significantly different between the three planets $\sim$4 Ga: the runaway greenhouse on Venus, the carbon-silicate cycle and ocean formation on Earth, and the CO$_2$-ice and H$_2$O-ice formation on Mars. 
Consequences of element partitioning for the atmospheric evolution  \textcolor{black}{during the late accretion onto} Venus, Earth, and Mars are investigated with a numerical model. 
We set upper limits to the partial pressures of CO$_2$ and H$_2$O on Earth and Mars, which corresponds to the state of phase equilibrium and carbon-silicate cycle. The final N$_2$ mass shrinks by $\sim$40\% and $\sim$15\% for Earth and Mars, respectively. 
The effect of element partitioning is found to be insufficient to reproduce the gaps.
For Venus, the survival of the primordial atmosphere through the late accretion may partially account for the present-day atmosphere.
Whereas on Mars, the atmospheric escape due to solar extreme UV and wind may have also influenced the atmospheric evolution.

\end{abstract}


\end{frontmatter}

%\linenumbers

\section{Introduction}\label{sec:introduction}
Atmospheres on terrestrial planets are believed to form as a consequence of volatile delivery and atmospheric erosion by impacts of numerous asteroids and/or comets \cite[e.g.,][]{abematsui1985, meloshvickery1989, deNiem2012, schlichting2015}.  The impacts do not only supply volatiles, but also remove part of the pre-existing atmospheres by impact erosion.\par

During the terrestrial planet formation, several tens of Mars-sized protoplanets were formed through the accretion of planetesimals. Subsequently, several giant impacts between these protoplanets occurred at the late stage of the terrestrial planet formation \cite[e.g.,][]{kokuboida1998}.  \textcolor{black}{If the protoplanets grew in the presence of the nebula, they have captured a primary atmosphere of nebular gas \cite[e.g.,][]{ikoma2006}.}\par

\textcolor{black}{According to the record on the Moon, the planets also experienced accretion of numerous impacts after the giant impact stage}, which is called late accretion \cite[e.g.,][]{bottke2010, chou1978}.  \textcolor{black}{Volatiles in the accretion bodies were vaporized forming a secondary atmosphere from impact degassing and/or volcanic degassing.} The late accretion includes the Late Heavy Bombardment (LHB), which is a spike of impact frequency between 4.1 to 3.8 Ga inferred from the craters on the Moon. 
\textcolor{black}{The excess of highly siderophile elements in Earth's mantle may reflect the additional meteoritic influx after core formation, which is called Late Veneer \cite[e.g.,][]{chou1978}. The total mass of the late accretion is estimated to range from $\sim$1\%  to $\sim$2.5\% of the planetary masses \cite[e.g.,][]{bottke2010, marchi2018}.
The period of the late accretion extends from the time of core-mantle differentiation, namely, the magma ocean phase, to that of the LHB. The magma ocean phase would last a few Myrs for Earth and Mars \cite[][]{elkins2008,hamano2013}.
}
The origin of the late accretion impactors is unknown, but geochemical studies \cite[e.g.,][]{fischer2017, dauphas2017} have shown that the isotopic compositions of the late veneer in Earth's mantle is more similar to enstatite chondrites than carbonaceous or ordinary chondrites.\par
 
Noble gases and nitrogen (N) contained in the atmosphere provide important clues to the origins of volatiles forming the atmosphere, oceans, and life (H, C, N, O, S, and P). Due to noble gases  \textcolor{black}{and to a lesser extent N} being chemically inert  \textcolor{black}{\cite[e.g.,][]{marty2012}}, they have mainly been partitioned into the atmosphere since $\sim$4 Ga, and so their abundances record the history of atmospheric formation and evolution. \par

\begin{figure}[htbp]
 \centering
 \includegraphics[width=8cm,clip]{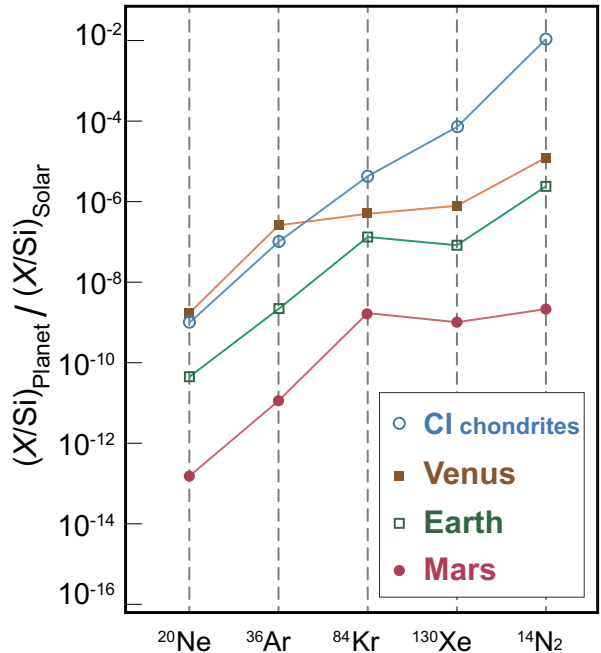}
 \caption{Abundances of noble gases and N in  \textcolor{black}{CI chondrites and} terrestrial planet atmospheres and known crustal reservoirs relative to Si with respect to the corresponding solar ratios (data are from \cite{pepin1991}).}
 \label{fig:noblegases_eps}
\end{figure}

\textcolor{black}{A similarity in the abundances of noble gases and N in the atmospheres of Venus, Earth, and Mars with chondrites may suggest the same origin \cite[e.g.,][]{pepin1991,marty2016}. However, there }are distinct gaps in the abundances in the atmospheres on \textcolor{black}{the three planets}: compared to Earth, Venus is enriched and Mars is depleted in noble gases by roughly two orders of magnitude, respectively (Figure \ref{fig:noblegases_eps}). The origin of these gaps is poorly understood. Since the noble gases and N are inert molecules, it is possible  \textcolor{black}{that} these gaps were created during the early stage of atmospheric formation and evolution.\par

 \textcolor{black}{
To a first approximation, the depletion pattern of noble gases and N in the planetary atmospheres follows the trend in chondritic values, but some differences between the planetary and chondritic abundances imply secondary effects: for instance, cometary additions to noble gases in the planetary atmospheres \cite[e.g.,][]{marty2017} and preferential loss of xenon \cite[][]{pujol2011}. 
The ratio of N to noble gases differs between three planets, which may be caused by the difference in partitioning of N into the mantle between them due to \textcolor{black}{N being less inert than} noble gases \cite[][]{mikhail2014,wordsworth2016}.
Whereas these differences suggest additional complexities of the atmospheric evolution, we aim to explain the most distinct gaps: the orders-of-magnitude gaps in the abundances noble gases and N between three planets.}\par
 \textcolor{black}{We investigate a possible explanation of these gaps involving the direct effect of impacts.}
\cite{gendaabe2005} proposed one possible scenario that explains the differences in the atmospheric contents of argon, krypton and xenon on Venus and Earth. By using numerical simulations, they demonstrated that the presence of oceans significantly enhances the loss of atmosphere during a giant impact and it can determine its subsequent atmospheric amount and composition. Protoplanets on orbits similar to Earth are expected to have had oceans, whereas those with Venus-like orbits are in the runaway greenhouse state  \textcolor{black}{\cite[e.g.,][]{hamano2013, hamano2015}}. As a result, a noble gas-rich primordial atmosphere survived on Venus, but not on Earth. However, the following late accretion might have influenced their atmospheres.\par

 \textcolor{black}{Even after the giant impact} stage, the terrestrial planets have experienced numerous impacts. \cite{deNiem2012} demonstrated that the atmospheric pressure had strongly increased for both Earth and Mars during the LHB.  \textcolor{black}{They also found that the initial pressure does not matter much compared to volatiles added by impact delivery.} This means that the differences in the atmospheric content on Venus and Earth might have changed after giant impacts. Therefore, it is important to investigate the atmospheric evolution during the late accretion to understand the origin and evolution of planetary atmospheres. In this study, we constructed a numerical model of impact degassing  \textcolor{black}{from impactors} and atmospheric erosion on Venus, Earth, and Mars during the late accretion.
Partitioning of elements in different surface reservoirs might influence the resulting noble gases and N abundances: the runaway greenhouse on Venus \cite[e.g.,][]{kasting1988, hamano2013}, the carbon-silicate cycle on Earth \cite[e.g.,][]{walker1981}, and the CO$_2$-ice formation on Mars \cite[e.g.,][]{forget2013, nakamura2003}. Although noble gases  \textcolor{black}{and N} are mainly partitioned into the atmosphere, the distinct environments on the three planets may have created the differences in  \textcolor{black}{their} concentrations in the atmospheres, leading to the various escape rates of noble gases  \textcolor{black}{and N} due to impact erosion (see subsection \ref{sec:partitioning} for details). \par

The purpose of this work is to investigate the atmospheric evolution of Venus, Earth, and Mars during the late accretion considering the effect of element partitioning. We describe our model in Section \ref{sec:model}. Numerical results for the noble gases and N abundances in the atmospheres and their  \textcolor{black}{dependence} on parameters will be shown in Section \ref{sec:results}. We discuss the origin and evolution of the atmospheres of the three terrestrial planets in Section \ref{sec:discussion}, and summarize our conclusions in Section \ref{sec:conclusion}.

\section{Numerical Model}\label{sec:model}

\subsection{Variables}
Table \ref{tab:variables} lists the variables and their meanings in our model.

\begin{landscape}
\begin{table}[htb]
\centering
\caption{ A list of variables and their meanings in our model}
  \begin{tabular}{clcl}\hline 
   $D$ & impactor diameter &$ T $ & planet surface temperature \\
   $V$ & entry velocity of impactor & $H$ & scale height \\ 
   $M_{\rm imp}$ & impactor mass & $\rho_0$ & atmospheric density \\ 
   $\rho_{\rm imp}$ & impactor density & $m_{\rm atm}$ & atmospheric mass \\
   $\Sigma_{\rm imp}$ & cumulative impactor mass & $m_{\rm atm,i}$ &mass of volatile species $i$ in the atmosphere \\
   $\Sigma{\rm _{imp}^{tot}}$ & total impactor mass & $\bar{m}$ & mean molecular mass \\
   $x$ & abundance of volatiles in an impactor &$m_i$ & volatile component molecular mass\\
   $x_i$ & abundances of volatile species $i$ in an impactor &$N_i$ & molecular numbers\\
   $X_{\rm C}$ & parameter for volatile abundance in an impactor & $P$ & total pressure\\
   $R_{\rm t}$ & planetary radius & $P_i$ & partial pressure\\
   $M_{\rm t}$ & planetary mass & $P^{\rm crit}_i$ & upper limit of partial pressure\\
   $u_{\rm esc}$ & planetary escape velocity & $\xi$ & impact energy parameter \\
   $\rho_{\rm t}$ &  \textcolor{black}{planetary mean density} & $\eta$ & atmospheric erosion efficiency \\ 
   $\alpha$ & impact angle & $\zeta$ & impactor's escaping efficiency\\
   $V_{\rm ground}$ & impact speed at ground level & $m_{\rm a}$ & eroded atmospheric mass \\
   $V_{\infty}$ & expansion speed of vapor plume & $m_{\rm v}$ & eroded impactor vapor mass \\ 
   \hline
  \end{tabular}
  \label{tab:variables}
\end{table}
\end{landscape}

In our model, the atmosphere is assumed to contain three components of volatiles: CO$_2$, H$_2$O, and N$_2$+non-radiogenic noble gas. The number $i$ represents each atmospheric component (1: CO$_2$, 2: H$_2$O, 3: N$_2$+noble gases). Since the amount of noble gases is small, we  \textcolor{black}{assume} that N$_2$ represents the third component.
 \textcolor{black}{We note that N can be more easily partitioned into other reservoirs than noble gases \cite[e.g.,][]{pepin1991}.
Because the difference in N partitioning between three planets are poorly constrained and controversial \cite[e.g.,][]{mikhail2014,wordsworth2016}, we assume that N is partitioned into the atmosphere.
The partitioning may decrease the N abundance in the atmosphere by a factor, but it would hardly change our discussion about the orders-of-magnitude difference in the abundances of noble gases and N.
} 

\subsection{Overview}
 \textcolor{black}{
We focus on the impact of element partitioning on the abundances of noble gases and N remaining after the late accretion in order to reconcile the gaps in the present-day atmospheres on Venus, Earth, and Mars. 
The partitioning CO$_2$ and H$_2$O into reservoirs other than the atmosphere would increase the relative concentration of noble gases and N in the atmosphere so that their erosion rate due to impacts might be enhanced.
}\par

Our model  \textcolor{black}{calculates} the amount of noble gases and N acquired by the terrestrial planets through impact degassing and atmospheric erosion during the late accretion  \textcolor{black}{\cite[e.g.,][]{pham2009, pham2011, pham2016}}.  \textcolor{black}{We note that we define impact degassing as the degassing from impactors. Neither degassing from target due to impacts nor volcanic degassing are considered in our model, but they will be discussed in subsection \ref{sec:volcanic}.}  \textcolor{black}{We assume all volatiles contained in the impactors were degassed during the impact, contributing to atmosphere delivery.} The assumption is valid for impactors whose impact velocities exceed ~5 km s$^{-1}$, which occurs for planets heavier than one Mars mass \cite[][]{catling2017}.  \textcolor{black}{We note that it is not strictly the case, as part of the impactor does not melt or is lost back to space \cite[][]{svetsov2015}.}  \textcolor{black}{After the impact, a huge jet, called a vapor plume, may rise from impact location and erode part of the atmosphere.} Some realistic numerical simulations of the atmospheric erosion have been performed by \cite{shuvalov2009} and \cite{svetsov2000, svetsov2007}, and we  \textcolor{black}{adopt} scaling laws obtained from these numerical simulations, as shown in subsection \ref{sec:atmosphericerosion}. \par

The net change in atmospheric mass per unit impactor mass is given by deterministic differential equations, 
\begin{eqnarray}
\frac{\mathrm{d}m_{\rm atm}}{\mathrm{d}\Sigma_{\rm imp}}&=&(1-\zeta)x-\eta 
\label{eq:model_AE} \\
\frac{\mathrm{d}(m_iN_i)}{\mathrm{d}\Sigma_{\rm imp}}&=&(1-\zeta)x_i-\eta\frac{(m_iN_i)}{m_{\rm atm}} ,
\label{eq:model_AEi}
\end{eqnarray}
where $m_{\rm atm}$ is the atmospheric mass, $\Sigma_{\rm imp}$ is the cumulative impactor mass, $x\ {\rm and}\ x_i$ are the abundance of all volatiles and that of species $i$, $m_i$ is the molecular mass, $N_i$ is the molecular number of each volatile component, $\eta$ is the erosion efficiency of the atmosphere, and $\zeta$ is that of the impactor vapor. In Equations \ref{eq:model_AE} and \ref{eq:model_AEi}, the first term on the right-hand side corresponds to the atmospheric supply and the second term to the atmospheric loss. The atmospheric erosion efficiency $\eta,$ and impactor's escaping efficiency $\zeta$ are defined as the masses of the removed atmosphere and impactor per unit impactor mass, respectively \cite[][]{shuvalov2009, svetsov2000, svetsov2007}.  \textcolor{black}{In the right-hand side of Equation \ref{eq:model_AEi}, atmospheric loss mass is proportional to the abundance of each species in the atmosphere. Therefore, element partitioning on the planet surface is important for atmospheric erosion.} We  \textcolor{black}{assume} an isothermal atmosphere. We also  \textcolor{black}{calculate} the total pressure and each partial pressure by,
\begin{eqnarray}
P&=&\frac{gm_{\rm atm}}{4\pi R_{\rm t}^2}\\
P_i&=&\frac{g\bar{m}N_i}{4\pi R_{\rm t}^2},
\label{eq:appendix_dPdz_int2}
\end{eqnarray}
where $P$ is the total pressure, $P_i$ is the partial pressures, $g$ is the gravitational acceleration, $R_{\rm t}$ is the planetary radius,  \textcolor{black}{and} $\bar{m}$ is the mean molecular mass of the atmosphere.\par

 \textcolor{black}{
Our model does not explicitly treat other mechanisms which might be related to the atmospheric evolution: survival of a primordial atmosphere \cite[][]{gendaabe2005}, atmospheric escape due to the solar extreme UV and wind \cite[e.g.,][]{jakosky2001, jakosky2017}, and volcanic degassing \cite[e.g.,][]{grott2011}.
The influence of these processes will be addressed in Discussion.}

In our model, we  \textcolor{black}{consider} the differences in element partitioning on the surfaces of three planets (subsection \ref{sec:partitioning}). We  \textcolor{black}{use} the present-day masses and sizes of Venus, Earth, and Mars and we  \textcolor{black}{assume} the amount of  \textcolor{black}{the} late accretion to be 1\% of each planetary mass \cite[][]{bottke2010}. Details of this model are given in the following sections.

\subsection{Elemental partitioning on planetary surfaces}\label{sec:partitioning}

\begin{figure}[htbp]
 \centering
 \includegraphics[width=8cm,clip]{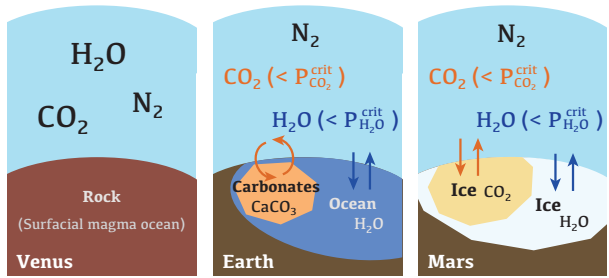}
 \caption{Surfacial environments on early Venus, Earth, and Mars. We  \textcolor{black}{assume} the runaway greenhouse state on Venus, formation of oceans and carbonate on Earth, and formation of H$_2$O-ice and CO$_2$-ice on Mars. The critical partial pressure $P^{\rm crit}_i$ set an upper limit to the partial pressure of chemical species $i$ due to phase equilibrium (H$_2$O on Earth and both H$_2$O and CO$_2$ on Mars) or global carbon-silicate cycle (CO$_2$ on Earth).}
 \label{fig:ElementsPartitioning}
\end{figure}

 \textcolor{black}{In order to consider partitioning of elements on the surface of each planet, we set the upper limits to the partial pressures} of H$_2$O and CO$_2$ considering the phase equilibrium and the steady state of the carbon-silicate cycle on Earth and Mars (Figure \ref{fig:ElementsPartitioning}). \par

The runaway greenhouse state \cite[][]{kasting1988}  \textcolor{black}{is} assumed on early Venus. 
All volatiles (H$_2$O, CO$_2$, and N$_2$)  \textcolor{black}{are} assumed to be partitioned into the atmosphere. 
The greenhouse effect of water vapor resulted in a high temperature and here we  \textcolor{black}{assume} $T = 2500\ {\rm K}$ \cite[][]{hamano2013},  \textcolor{black}{though early Venus in a habitable condition has been also proposed \cite[][]{way2016}}. 
 \textcolor{black}{In fact, Venus might have been heated by repeated large impacts in a short time. Subsequently, the surface temperature would have decreased quite fast and stay below 2000 K. We will discuss the dependency on the assumed temperature in subsection \ref{sec:TandPcrit}.} \par
 
The formation of oceans and a carbon-silicate cycle  \textcolor{black}{are} assumed on early Earth. The surface temperature on Earth is estimated to be 273 K to 360 K \cite[][]{kasting1993} and so we  \textcolor{black}{assume} $T$ = 288 K. The vapor-liquid equilibrium  \textcolor{black}{sets} an upper limit to the partial pressure of H$_2$O, $P^{\rm crit}_{\rm H_2O} = 1.7\times10^{-2}\ {\rm bar}$ \cite[][]{murphykoop2005}. The negative feedback of the carbon-silicate cycle forces $P_{\rm CO_2}$ to reach the steady state. We  \textcolor{black}{treat} the effect of carbon-silicate cycle by simply setting an upper limit to $P_{\rm CO_2}$  \textcolor{black}{and treat it as a free parameter. We assume a value} expected as the steady state $P^{\rm crit}_{\rm CO_2} = 1$ bar in the standard model and we also  \textcolor{black}{investigate} the dependencies on its value by considering the parameter space from 0.1 bar to 10 bar \cite[][]{kasting1993}. We note that this assumption  \textcolor{black}{neglects} the time lag to reach the steady state, which maximizes the effect of the difference in the surfacial environments on the abundances of noble gases and N$_2$.\par

 \textcolor{black}{H$_2$O-ice and CO$_2$-ice are assumed to have formed on early Mars. Possible surface temperatures on early Mars range from 200 K \cite[][]{forget2013} to 273 K, which allows an ocean to form. We choose T = 223 K as the reference value.} The vapor-ice equilibrium  \textcolor{black}{sets} upper limits to $P_{\rm H_2O}$ and $P_{\rm CO_2}$. We  \textcolor{black}{assume} $P^{\rm crit}_{\rm H_2O} = 3.9\times 10^{-6}\ {\rm bar} $ \cite[][]{murphykoop2005}. The upper limit of CO$_2$ partial pressure is assumed to be $P^{\rm crit}_{\rm CO_2} = 3$ bar in the standard model, which was obtained as a maximum pressure in the condition without CO$_2$-ice by a 3D global circulation model of a CO$_2$-dominated atmosphere \cite[][]{forget2013}. We also  \textcolor{black}{treat} $P^{\rm crit}_{\rm CO_2}$ as a parameter and  \textcolor{black}{calculate} for the range from 6 mbar to 3 bar in the calculation.

\textcolor{black}{
The period of the late accretion extends from the magma ocean phase to that of the LHB. We neglect interactions between the atmospheres and the magma oceans in our model. The presence of magma oceans might have allowed volatiles to also be partitioned into them \cite[e.g.,][]{moore1995,abe1988,zahnle1988,elkins2008,salvador2017,marcq2017}. 
Calculating the evolution of the magma ocean depth is required to estimate the partitioning into magma oceans. 
The amount of dissolved H$_2$O can be larger than that in the atmosphere for a deep magma ocean, but smaller for a shallow magma ocean \cite[][]{hamano2013}.
The partitioning into a magma ocean is less efficient for CO$_2$ and N$_2$ \cite[][]{salvador2017}.
In contrast to Venus, the magma ocean phase would have been short (a few Myrs) for Earth and Mars \cite[][]{elkins2008,hamano2013}.
In addition, a large impact itself could induce large scale melting and degassing of the target \cite[][]{gillmann2016}.
Instead of constructing a complex model taking the surface evolution into account, we decide to fix the surface conditions of three planets in the period of the late accretion, but we evaluate the influence of the difference in the element partitioning for wide parameter ranges, which cover the surface environments evolved through time.
This approach allows us to infer a limit on the possible influence of evolving surface conditions.}\par

\subsection{Late accretion  \textcolor{black}{impactors}}\label{sec:impactor}
The origin of late accretion impactors is unknown and so we  \textcolor{black}{treat} the abundance of volatiles in an impactor as a parameter $X_{\rm C}$, assuming the volatile abundances of carbonaceous chondrite \textcolor{black}{-like volatile-rich impactors} as a reference value. We  \textcolor{black}{assume} the reference abundances are as follows and multiplied the parameter $X_{\rm C}$  \textcolor{black}{in order to simplify the model}: an impactor contains 2\% CO$_2$ \cite[][]{deNiem2012}, 10\% H$_2$O \cite[][]{raymond2004}, and 0.2\% N$_2$ \cite[][]{gradywright2003}. \par

Recent geochemical studies suggested that the late accretion was mainly composed of enstatite-chondrite-like, relatively volatile-poor impactors \cite[][]{fischer2017, dauphas2017}. Therefore, in most of our calculations, we  \textcolor{black}{assume} $X_{\rm C}=0.1$ as a nominal value.  \textcolor{black}{Since the composition of enstatite chondrites have a large variation, we simply vary the parameter value as $X_{\rm C}=1, 0.1,$ and 0.01.}
The total mass of  \textcolor{black}{impactors} $\Sigma{\rm _{imp}^{tot}}$  \textcolor{black}{is} assumed to be 1\% of each planetary mass.\par

The initial abundance of the atmosphere and its composition at the end of the giant impact stage are also unknown factors. \cite{deNiem2012} argued that the initial atmospheric pressure is only an additive constant to the final pressure. In our standard model, we  \textcolor{black}{assume} the initial atmospheric surface pressure 0.1 bar and its composition are the same as the volatile components in the impactor. The dependence on initial atmospheric surface pressure is also discussed in subsection \ref{sec:initialpressure}.

\subsection{ \textcolor{black}{Atmospheric erosion and impactor retention models}}\label{sec:atmosphericerosion}

When a small body enters the atmosphere of a planet, the atmospheric gas is compressed and displaced, decreasing the impactor speed \cite[e.g.,][]{svetsov2000, svetsov2007}. The impactor's kinetic energy converts into the kinetic energy of gas flow expanding outward \cite[e.g.,][]{svetsov2000}, forming a vapor plume of partially vaporized impactor and target planetary material \cite[e.g.,][]{svetsov2007, shuvalov2009}. Through this sequence of an impact, part of the impact vapor plume and atmospheric gas is removed.  
\cite{pham2016} developed a tangent \textcolor{black}{plane} model with an upper limit on the amount of eroded atmospheric gas, which is not included in our model. \textcolor{black}{We note that \cite{svetsov2005} demonstrated hydrodynamical simulations of large impacts over hundreds of kilometers in diameter regarding the impact-produced rock vapor atmosphere. }
\par

Realistic numerical simulations of atmospheric erosion were given in 3D geometry by \cite{shuvalov2009}, and in 2D cylindrical geometry by \cite{svetsov2007}. Both studies suggested scaling formulas for atmospheric effects.  \textcolor{black}{\cite{shuvalov2014} later modified his previous model \cite[][]{shuvalov2009} by considering the effect of aerial burst and fragmentations of projectile during the entry of small projectile into a dense atmosphere, which may slightly enhance the atmospheric erosion for the case of Venus compared to \cite{shuvalov2009}. } \cite{deNiem2012} pointed out that the atmospheric erosion behavior of the \cite{shuvalov2009} model is uncertain for large impactors because he only used diameters from 1 km to 30 km for the impactor. 
Therefore, we  \textcolor{black}{adopt} \cite{svetsov2000, svetsov2007} for the atmospheric erosion efficiency $\eta$, and \cite{shuvalov2009} for the impactor's escaping efficiency $\zeta$.\par

\subsubsection{Svetsov (2000, 2007) model}
\cite{svetsov2000} proposed an analytical model with emphasis on the fragmentation of objects decelerated by the atmosphere.
 \textcolor{black}{He} regarded the meteor entry phase in an analytical pancake model \cite[][]{hills1993}.
The atmospheric erosion efficiency is given by, 
\begin{eqnarray}
\eta{\rm_{Sv}}&\equiv& \frac{m_{\rm a}}{M_{\rm imp}}\nonumber\\
&=&\frac{3\rho_{\rm 0}}{4\rho_{\rm imp}}\left(\frac{2H}D+\frac{16H^2\rho_0^{\frac12}}{3D^2\rho_{\rm imp}^{\frac12}}+\frac{16H^3\rho_{\rm 0}}{D^3\rho_{\rm imp}}\right)
\cdot \left(\frac{\int^1_X(1-x^2)^{k}dx}{\int^1_0(1-x^2)^{k}dx}\right)\cdot f_\alpha,
\label{eq:Sv00_Ea_define}
\end{eqnarray}
where $m_{\rm a}$ is eroded atmospheric mass, $\rho_0$ is atmospheric density, $\rho_{\rm imp}$ is impactor density, $H$ is scale height of the planetary atmosphere, $D$ is impactor diameter, $X$ is defined by $X\equiv \min\{1, u_{\rm esc} /V_\infty\}$: the ratio of the planetary escape velocity $u_{\rm esc}$ to the asymptotic expansion speed of the vapor plume $V_\infty = V_{\rm ground}\sqrt{4\gamma/(\gamma-1)}$, $\gamma$ is the adiabatic coefficient of the gas, and $k$ indicates how the gas expands outward related to $\gamma$. We  \textcolor{black}{use} a value of $\rho_{\rm imp}=3.32\ {\rm g/cm^3}$ suggested by \cite{shuvalov2009}, and $k=5$ corresponding $\gamma=\frac{13}{11}$ suggested by \cite{svetsov2007}. The impact speed at ground level $V_{\rm ground}$ is related to the entry velocity $V$ as,
\begin{eqnarray}
V_{\rm ground} = V \exp \left( - \frac{\rho_{\rm 0}}{\rho_{\rm imp}}\left(\frac HD + \frac{2H^2\rho_{\rm 0}^{\frac12}}{\frac34 D^2\rho_{\rm imp}^{\frac12}}\right) \right).
\label{eq:Sv00_inputparameter}
\end{eqnarray}
In the calculation, the impactor velocity distribution  \textcolor{black}{is} considered for the entry velocity $V$ as explained in subsection \ref{sec:numericalmethod}. \par

The last term $f_\alpha$ in Equation \ref{eq:Sv00_Ea_define} is a correction factor for the impact angle distribution by \cite{svetsov2007} that is defined as, 
\begin{eqnarray}
f_\alpha &=& \int_0^{\frac \pi 2}d\alpha\left(1+2\sin(2\alpha)\right)^2\sin(2\alpha)\nonumber\\
&=&5+\pi-\frac43,
\label{eq:Sv07_falpha}
\end{eqnarray}
where $\alpha$ is impact angle and is applied in our model.

\subsubsection{Shuvalov (2009) model}
Our model  \textcolor{black}{calculates} the escaping efficiency of the impactor vapor plume by using the model of \cite{shuvalov2009}. We note that using a different model  \textcolor{black}{\cite[][]{vickerymelosh1990}} showed similar results (less than 10 \% difference in the final N$_2$ mass at the point of $\Sigma_{\rm imp}=0.01\times M_{\rm t}$).

\cite{shuvalov2009} is the most extensive 3D hydrocode study that investigated the atmospheric erosion and impactor retention using realistic equations of state. Oblique impacts were also considered in this model. The dimensionless mass of escaping projectile material $\zeta$ was approximated as a simple analytical formula with the dimensionless erosion power $\xi$ by, 
\begin{eqnarray}
\zeta{\rm_{Sh}}&=&\frac{m_{\rm v}}{M_{\rm imp}}\nonumber\\&=&\min \left\{0.035\frac{\rho{\rm_t}}{\rho_{\rm imp}}\cdot\frac V{u_{\rm esc}}(\log \xi-1)\ ;0.07\frac{\rho_{\rm t}}{\rho_{\rm imp}}\cdot\frac V{u_{\rm esc}};\ 1\right\}
\label{eq:model_zeta}\\
\xi&=&\left(\frac DH\right)^3\frac{\rho_{\rm imp}\rho_{\rm t}}{\rho_{\rm 0}(\rho_{\rm imp}+\rho_{\rm t})}\max\left\{\frac{V^2}{u_{\rm esc}^2}-1 ; 0\right\}, 
\label{eq:Sh09_x}
\end{eqnarray}
where $m_{\rm v}$ is eroded impactor vapor mass, $\rho_{\rm t}$ is planetary  \textcolor{black}{material} density and the value of $\rho_{\rm t}=2.63\ {\rm g/cm^3}$ suggested in \cite{shuvalov2009}  \textcolor{black}{is} also adopted in our model.\par

By substituting erosion efficiencies $\eta_{\rm Sv},\ \zeta_{\rm Sh}$ into Equation \ref{eq:model_AE} and \ref{eq:model_AEi}, we  \textcolor{black}{calculate} the evolution of early atmospheres during the late accretion.

\subsection{Numerical method}\label{sec:numericalmethod}
We numerically  \textcolor{black}{integrate} Equations \ref{eq:model_AE} and \ref{eq:model_AEi} over time until the cumulative impactor mass reached the total mass: 1\% of the planetary mass.
At each numerical step of the cumulative impactor mass, the erosion efficiencies $\eta$ and $\zeta$  \textcolor{black}{are} averaged over the impactor size and velocity distributions by, 
\begin{eqnarray}
\bar{\eta}=\int_M\int_V \eta f_V(V)f_M(D) \mathrm{d}V \mathrm{d}D\\
\bar{\zeta}=\int_M\int_V \zeta f_V(V)f_M(D) \mathrm{d}V \mathrm{d}D,
\label{eq:model_average}
\end{eqnarray}
where $ f_V(V)$ is the frequency distribution function of the impact velocity $V$ and $f_M(D)$ is that of impactor size $D$, which is a mass-weighted frequency distribution function defined as follows.\par

The impactor size distribution  \textcolor{black}{is} assumed to depend on the impact  \textcolor{black}{diameter} as $ \mathrm{d}N(D)/\mathrm{d}D\propto D^{-3}$ where $N(D)$ is the number of objects of diameter smaller than $D$, which corresponds to that of the present-day main belt asteroids \cite[][]{bottke2005}. The average erosion efficiency over the projectile size range of $10^{-1.5}\ {\rm km}\leq D\leq10^3\ {\rm km}$  \textcolor{black}{is} used in our model. By considering the impactor volume, the probability function of impactor mass  \textcolor{black}{is} obtained as,
\begin{eqnarray}
f_M(M)\mathrm{d}M&\propto& D^3\cdot\frac{\mathrm{d}N(D)}{\mathrm{d}D}\cdot\mathrm{d}D\nonumber\\
&\propto& D^0 dD\nonumber\\
&\propto& 10^{\log_{10}D}\mathrm{d}(\log_{10}D)
\label{eq:dMdD}
\end{eqnarray}
where the integration of $D$  \textcolor{black}{is} separated into logarithmic bins.\par

For the impact velocity distribution, we  \textcolor{black}{assume} a Rayleigh distribution, which corresponds to the Gaussian eccentricity impactors from the terrestrial planet feeding zone \textcolor{black}{. T}his assumption  \textcolor{black}{is based on} \cite{idamakino1992}, which demonstrated that the eccentricity of planetesimals in the feeding zone excited by a protoplanet can be approximated by Gaussian. The probability distribution function of the impact velocity  \textcolor{black}{is} obtained by,
\begin{eqnarray}
f_{V}(V)\propto\sqrt{V^2-u_{\rm esc}^2}\exp\left\{-\frac{V^2-u_{\rm esc}^2}{2u_{\rm esc}^2}\right\}, 
\label{eq:Raylie}
\end{eqnarray}
and we  \textcolor{black}{calculate} the average over the range of $10.4\ {\rm km\ s}^{-1}\leq V\leq 44.8\ {\rm km\ s}^{-1}$ for Venus, $11.2\ {\rm km\ s}^{-1}\leq V\leq 48.4\ {\rm km\ s}^{-1}$ for Earth, and $5.03\ {\rm km\ s}^{-1}\leq V\leq 21.8\ {\rm km\ s}^{-1}$ for Mars.  \textcolor{black}{We note that the dependence on the maximum value of the impactor velocity distribution is small}.\par

The fractional impactor mass increment $\Delta\Sigma_{\rm imp}$  \textcolor{black}{is} calculated at every step by,
\begin{eqnarray}
\Delta\Sigma_{\rm imp}=\Sigma_{\rm imp}^{(n+1)}-\Sigma_{\rm imp}^{(n)}=10^{-2}\cdot\frac{\delta m_{\rm atm}}{m_{\rm atm}^{(n)}}=\frac{10^{-2}}{(1-\zeta)x-\eta}\cdot m_{\rm atm}^{(n)},
\label{eq:model_AEstep}
\end{eqnarray}
where $n$ and $n+1$ are the numbers of steps.

\section{Results}\label{sec:results}

\subsection{The evolution of atmospheres during the late accretion}
To investigate the effect of element partitioning on the final noble gases and N abundances of planets  \textcolor{black}{after} the late accretion, we calculated the evolution of the masses and compositions of the atmospheres on the terrestrial planets. The partial pressure and mass of each component at every step were obtained by solving Equation \textcolor{black}{s} \ref{eq:model_AE} and \ref{eq:model_AEi}.\par

Table \ref{tab:parameters} summarizes the values of input parameters applied in the simulations for the three standard planet models according to model descriptions in Section \ref{sec:model}. The results in the standard models are shown in Figures 4--7.\par

\begin{table}[htb]
\centering\small
\caption{Input parameters for standard models}
  \begin{tabular}{lccc}\hline 
    & Venus model & Earth model & Mars model\\ \hline
    Planetary radius $R_{\rm t}$ & 6010 km & 6370 km & 3390 km\\
    Planetary mass $M_{\rm t}$ & 4.87$\times 10^{24}$ kg & 5.97$\times 10^{24}$ kg & 6.39$\times 10^{23}$ kg\\
    Surface temperature $T$ & 2500 K & 288 K & 223 K\\
    Pressure upper limit $P^{\rm crit}_{\rm CO_2}$ & none & 1 bar & 3 bar\\
    Pressure upper limit $P^{\rm crit}_{\rm H_2O}$ & none & $1.7\times 10^{-2}\ {\rm bar}$ & $3.9\times 10^{-6}\ {\rm bar}$\\ \hline
  \end{tabular}
  \label{tab:parameters}
\end{table}

\begin{figure}
 \centering
 \includegraphics[width=12cm,clip]{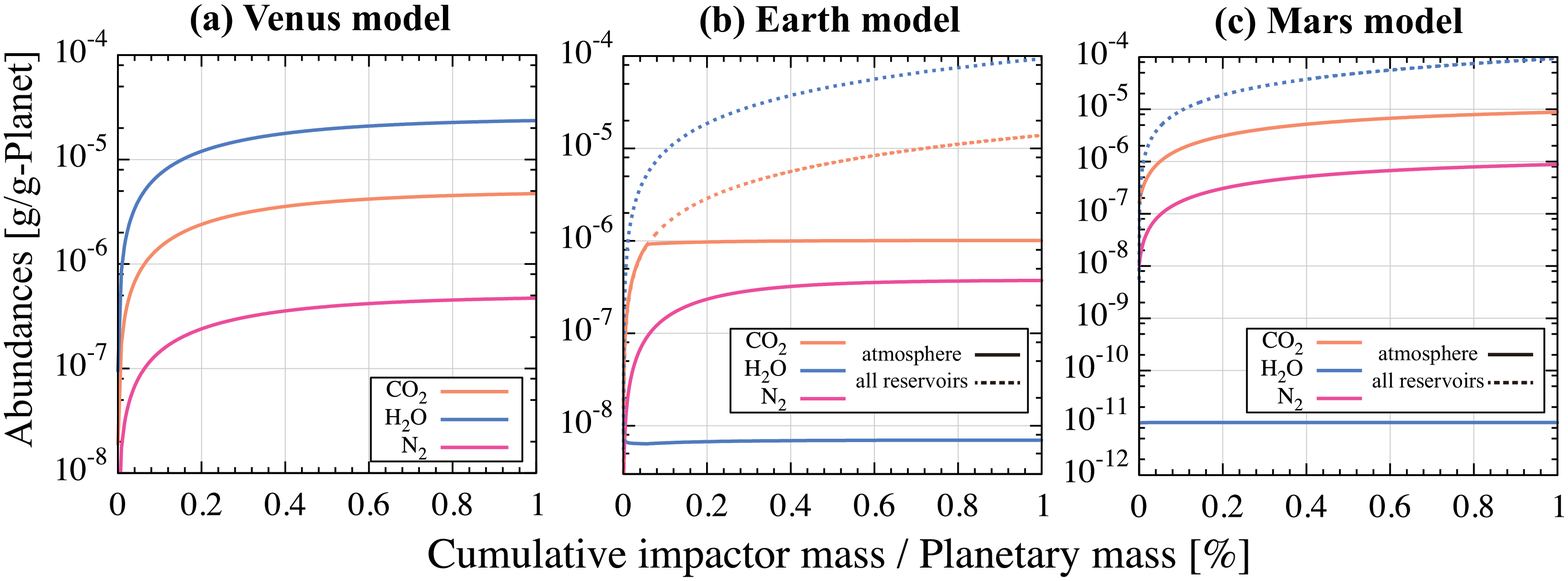}
 \caption{The evolution of  \textcolor{black}{the abundances of CO$_2$, H$_2$O, and N$_2$} on the early terrestrial planets during the late accretion as a function of the cumulative impactor mass hitting each planet.  \textcolor{black}{Solid lines correspond to the amount in the atmosphere and dashed lines correspond to the amount in all reservoirs on each planet. Plotted abundances are scaled by each planetary mass. }(a) The results of the standard Venus model. Colors represent each volatile component: CO$_2$, H$_2$O, and N$_2$. Surface temperature: $T=$ 2500 K, Impactor: Enstatite chondrite-like composition. Pressure limits: none. (b) The results of the standard Earth model. Surface temperature: $T=$ 288 K, Impactor: Enstatite chondrite-like composition. Pressure limits: $P^{\rm crit}_{\rm CO_2}=1$ bar, $P^{\rm crit}_{\rm H_2O}=1.7\times10^{-2}\ {\rm bar}$. (c) The results of the standard Mars model. Surface temperature: $T=$ 223 K, Impactor: Enstatite chondrite-like composition. Pressure limits: $P^{\rm crit}_{\rm CO_2}=3$ bar, $P^{\rm crit}_{\rm H_2O}=3.9\times10^{-6}\ {\rm bar}$. }
 \label{fig:stdVEMpressure} 
 \end{figure}

Figure \ref{fig:stdVEMpressure} shows the atmospheric evolution  \textcolor{black}{and the volatile abundances in the all reservoirs} on Venus, Earth, and Mars. 
The  \textcolor{black}{abundances} of three components during the late accretion are plotted as a function of the ratio of the cumulative impactor mass to the planetary mass. 
The initial atmospheric pressure of 0.1 bar was assumed. The partial pressures of particular components -- CO$_2$ on Earth and H$_2$O on Earth and Mars -- reached their upper limits, which are the critical upper limits by condensation, coagulation, and carbonate formation  \textcolor{black}{and so their abundances in the atmospheres stopped to increase}. It means that the phase equilibrium between gas and liquid or solid was attained for CO$_2$ and H$_2$O on Earth and H$_2$O on Mars. The other partial pressures  \textcolor{black}{and abundances in the atmosphere} increased rapidly at the early phase because the volatile supply dominated.  \textcolor{black}{They} subsequently reached a steady state. These steady states correspond to the balance between atmospheric supply and loss.  \textcolor{black}{The total abundances of each component are also plotted in Figure \ref{fig:stdVEMpressure}. Since the amounts of CO$_2$ and H$_2$O exceeded the upper limits of partial pressures, they started to be partitioned into the solid or liquid reservoirs so that large amounts of CO$_2$ and H$_2$O were preserved on Earth and H$_2$O on Mars.}\par

\begin{figure}
 \centering
 \includegraphics[width=7.5cm,clip]{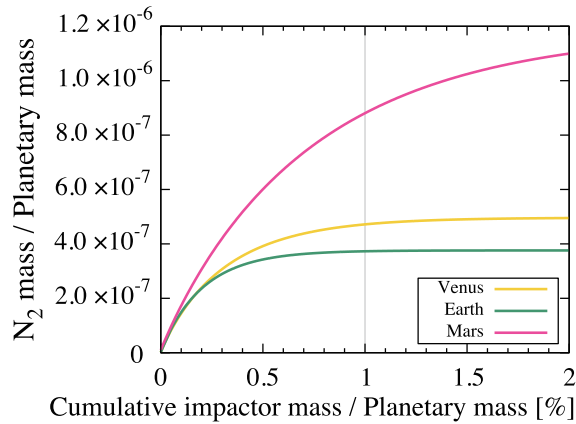}
 \caption{The evolution of N$_2$ mass of Venus, Earth, and Mars during the late accretion as a function of the cumulative impactor mass. 1\% of the planetary mass (gray vertical line) corresponds to the end of the late accretion. Colors represent each planet: Venus (yellow), Earth (green), and Mars (magenta). Both the plotted cumulative impactor mass and N$_2$ mass are divived by each planetary mass. }
 \label{fig:Std3PlanetsN2mass}
\end{figure}

Figure \ref{fig:Std3PlanetsN2mass} compares the evolution of N$_2$ mass on Venus, Earth, and Mars in the standard models. The ratio of N$_2$ mass to the planetary mass is plotted as a function of the ratio of the cumulative impactor mass to the planetary mass from 0\% to 2\%. 
The resulting N$_2$ mass (scaled by the planetary mass) of the Mars model is about twice as large as that of Venus and Earth. It suggests that Mars obtained more noble gases and N than Venus and Earth during the late accretion, which is inconsistent with present-day atmospheres (Figure \ref{fig:noblegases_eps}). This is because Mars' size and  \textcolor{black}{the} ratio of mass to surface area are smaller than those of Venus and Earth. We will discuss this point further in subsection \ref{sec:smallmars}. The N$_2$ mass of Venus obtained in the simulation is also larger than Earth, but the difference was less than twice as much.  \textcolor{black}{The loss of N$_2$ from Venus was suppressed by no element partitioning. However, \textcolor{black}{this effect was} compensated by the higher temperature and density of the atmosphere, both of which enhanced the atmospheric erosion.} The dependence on the surface temperature will be shown in subsection \ref{sec:TandPcrit}. The effect of the total atmospheric mass on the atmospheric erosion will be discussed in subsection \ref{sec:negativefeedback}.

\subsection{The effect of element partitioning}\label{sec:partitioningresult}

\begin{figure}[htbp]
 \centering
 \includegraphics[width=7.5cm,clip]{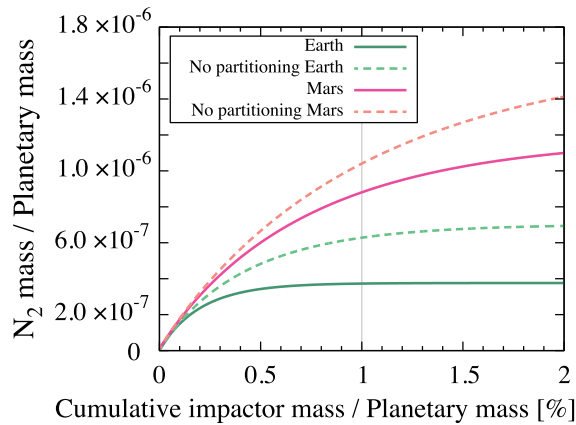}
 \caption{Effect of element partitioning on the final N$_2$ abundance for Earth and Mars. The magenta lines correspond to the evolution of N$_2$ mass on Mars and green lines are that on Earth. Solid lines correspond to the  \textcolor{black}{cases} with the critical partial pressures of CO$_2$ and H$_2$O, and the dashed lines show the cases without the upper limits.}
 \label{fig:StdvsContEarthMars}
\end{figure}

To investigate the effect of element partitioning on the abundances of noble gases and N acquired by terrestrial planets, we calculated the cases without upper limits of partial pressures for Earth and Mars. In other words, we assumed no partitioning of volatiles into liquid or solid phase on Earth and Mars. Figure \ref{fig:StdvsContEarthMars} compares the results of the Earth model and Mars model. In the model where the upper limits on the partial pressures of CO$_2$ and H$_2$O were assumed, the resulting N$_2$ mass is $\sim$40\% and $\sim$15\% smaller for Earth and Mars, respectively, than it is in the model where no upper limit was assumed. This result suggests that the abundances of noble gases and N on planets decreases due to element partitioning of other volatiles into liquid or solid phase during the late accretion.  \textcolor{black}{This is because the ratio of such inert gases in the atmosphere increased when some main components (CO$_2$ and H$_2$O) are hidden into other reservoirs. Because the erosion efficiency is proportional to the abundances of each species, a larger amount of inert gases was lost.} However, the differences of resulting N$_2$ masses are not sufficiently large to explain that of present-day atmospheres. 
One cause of the small effect of element partitioning is the dependence of the atmospheric erosion efficiency on the atmospheric mass, which will be shown in subsection \ref{sec:negativefeedback}.
The N$_2$ mass evolution also depends on free parameters such as the upper limits of partial pressures and the surface temperature. The dependences on these parameters will be shown in the next section.

\subsection{The dependences on temperature and upper limit of partial pressure}\label{sec:TandPcrit}

We changed the free parameters in our model over a range of realistic settings for the early terrestrial planets.
\begin{figure}
 \centering
 \includegraphics[width=12cm,clip]{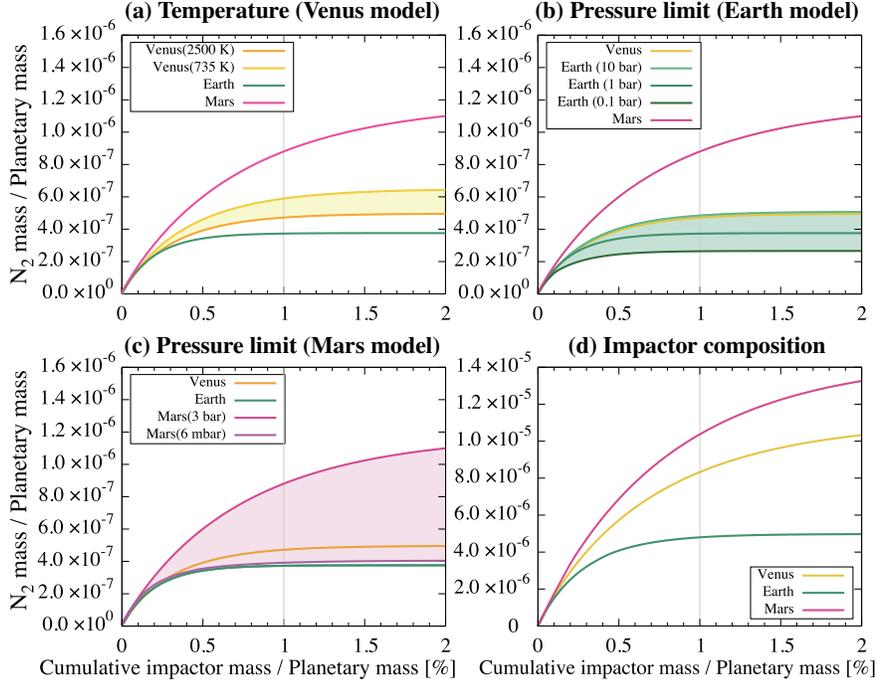}
 \caption{The evolutions of N$_2$ abundance as a function of cumulative impactor mass in the parameter research. (a) Dependence of final N$_2$ mass on surface temperature on Venus. The yellow area is for Venus model with the surface temperature from 735 K (yellow line) to 2500 K (orange line). N$_2$ mass evolution of Earth (green line) and Mars (magenta line) model are also plotted.  \textcolor{black}{(b) Dependence of final N$_2$ mass on the upper limit of CO$_2$ partial pressure on Earth. The green area represents the range of N$_2$ mass obtained in Earth model with the upper limit of CO$_2$ partial pressure $P_{\rm CO_2}^{\rm crit}$ ranging from 0.1 bar (dark green line) to 10 bar (light green line). N$_2$ mass evolution of nominal Venus (yellow line), Earth (green line, $P_{\rm CO_2}^{\rm crit}=1\ {\rm bar}$), and Mars (magenta line) models are also plotted. (c) Dependence of final N$_2$ mass on the upper limit of CO$_2$ partial pressure on Mars. The magenta area represents the range of N$_2$ mass obtained in Mars model with the upper limit of CO$_2$ partial pressure ranging from 6 mbar (purple line) to 3 bar (magenta line). N$_2$ mass evolution of Venus (yellow line) and Earth (green line) models are also plotted.} (d) Same as Figure \ref{fig:Std3PlanetsN2mass}, but for the case with the impactor containing volatiles by $X_{\rm C}=1$, which corresponds to the carbonaceous chondrite-like impactor. }
 \label{fig:ParameterResearch} 
\end{figure}
Figure \ref{fig:ParameterResearch}(a) shows the dependence of the N$_2$ mass on the surface temperature on Venus. We used values from 735 K, which is the present-day mean surface temperature on Venus \cite[][]{marov1978}, to 2500 K. The high temperature case resulted in less remaining N$_2$, but the difference between the resulting N$_2$ masses was less than 20\%. This is because the atmospheric erosion efficiency has a small positive dependence on the scale height $H$ and $H$ becomes higher with increased temperature. \par

Figures \ref{fig:ParameterResearch}(b) and \ref{fig:ParameterResearch}(c) shows the dependence of the N$_2$ mass on the upper limit of partial pressure on Earth and Mars. We changed the value of the upper limit of CO$_2$ partial pressure from 0.1 bar to 10 bar on Earth and from 3 bar to 6 mbar on Mars, where the latter corresponds to the mean atmospheric pressure on current Mars \cite[e.g.,][]{jakosky2001}.  \textcolor{black}{In the case where we assumed a smaller upper limit, a smaller N$_2$ mass was obtained.} When the upper limit of CO$_2$ partial pressure is smaller, the amount of CO$_2$ in the atmosphere diminishes and, consequently,  \textcolor{black}{the relative abundance of N$_2$ becomes larger}. The larger amount of N$_2$ would be removed by the atmospheric erosion by expanding vapor plumes (see subsection \ref{sec:smallmars}). The N$_2$ mass of the case setting $P^{\rm crit}_{\rm CO_2}=10$ bar on Earth was almost the same as that of the Venus model and the final amount of N$_2$ differed by twice as much within the parameter range for Earth. The N$_2$ mass on Mars obtained in the case of $P^{\rm crit}_{\rm CO_2}=6$ mbar was still larger than that of the Earth model. 
The differences between the final N$_2$ masses of the three planets are still small. The dependence on the upper limit of partial pressure turned out to be less important for the final amount of N$_2$. \par

These results showed that the effect of element partitioning on the early terrestrial planets is insufficient to reproduce the present-day atmospheres in spite of considering the large parameter ranges.
 \textcolor{black}{Though we neglected the evolution of surface environments through time, our parameter survey shown here suggests that the gaps in noble gases and N abundances is difficult to be reproduced even with the evolving surface conditions.} \par

\subsection{Impactor composition dependence}\label{sec:Cchond}
Figure \ref{fig:ParameterResearch}(d) shows the same scenario as given in Figure \ref{fig:Std3PlanetsN2mass}, but for the case with a volatile-rich impactor. We treated the volatile abundance in an impactor as a parameter by using the parameter X$_{\rm C}$ (subsection \ref{sec:impactor}). In this case for the carbonaceous chondrite-like composition, we assumed $X_{\rm C}=1$. Compared to Figure \ref{fig:Std3PlanetsN2mass}, the resulting N$_2$ mass on each planet increased by an order of magnitude, but the relative differences between Venus, Earth, and Mars were almost the same. We found that the difference in the final N$_2$ mass (scaled by the planetary mass) between Venus, Earth, and Mars is smaller than an order of magnitude regardless of the impactor composition. \par
\begin{figure}
 \centering
 \includegraphics[width=8cm,clip]{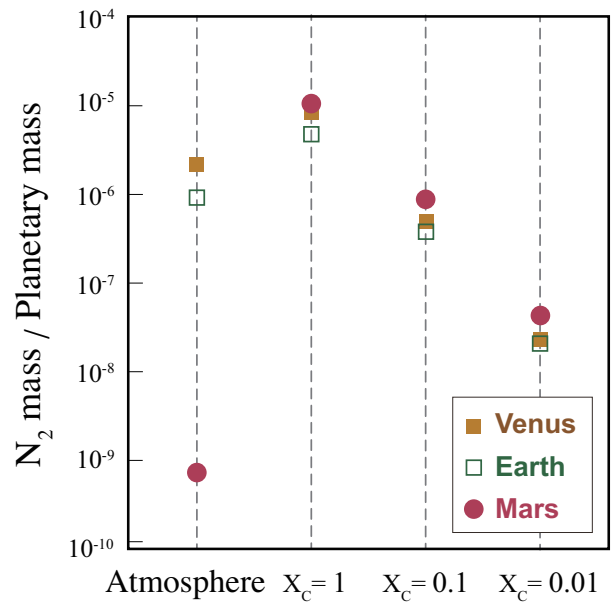}
 \caption{ \textcolor{black}{The final abundances of N after the late accretion ($\Sigma_{\rm imp}=0.01 M_{\rm t}$) in the cases assuming impactor composition of $X_{\rm C}=1$, 0.1, and 0.01. The abundances in the present-day atmosphere on each planet are also plotted \cite[][]{pepin1991}. The plotted N$_2$ mass is scaled by each planetary mass. We note that the abundances of noble gases are assumed to vary in proportion to those of N.}}
 \label{fig:ImpactorCompositionResults_Nplanet} 
\end{figure}

 \textcolor{black}{We also calculated the evolution for the case $X_{\rm C}=0.01$, where impactor contains a much smaller amount of volatiles than the nominal model. The results of these calculations and the abundances of N in the present-day atmospheres are compared in Figure \ref{fig:ImpactorCompositionResults_Nplanet}. Though the resulting N$_2$ mass varied by orders of magnitude depending on the assumed $X_{\rm C}$ values, we found that the difference in N$_2$ masses scaled by planetary masses between three planets is too small to reproduce the difference between present-day Venus, Earth, and Mars in all cases.}

\section{Discussion}\label{sec:discussion}
\subsection{Analysis of the difference in N$_2$ mass in the numerical model}\label{sec:smallmars}
The effect of element partitioning was shown in subsection \ref{sec:partitioningresult}. 
Condensation, coagulation, and carbonate formation of CO$_2$ and H$_2$O induced a higher concentration of noble gases and N in the atmosphere. 
Subsequently, a larger amount of noble gases and N was eroded. 
In the model where the lower upper limit of partial pressure was assumed, this effect becomes more efficient because of the higher concentration of N$_2$ (see subsection \ref{sec:TandPcrit}). \par

However, the effect of element partitioning on atmospheric evolution was found to be insufficient to explain the differences in abundances of noble gases and N in the present-day atmospheres on Venus, Earth, and Mars. 
The resulting N$_2$ mass on Mars even exceeded that on Venus in spite of the assumption of the upper limits on partial pressures.\par

First, we show how the resulting N$_2$ mass should differ between the three planets considering their differences in size without the effect of element partitioning. The N$_2$ mass scaled by the planetary mass can be  \textcolor{black}{transformed} as, 
 \begin{eqnarray}
\frac{m_{\rm atm, N_2}}{M_{\rm t}}=\frac{4\pi R_{\rm t}^2H\rho_0X_{\rm N_2}}{M_{\rm t}}
\propto \frac{\rho_0}{R_{\rm t}^2\rho_{\rm t}^2},
\label{eq:N2massPlanetmass}
\end{eqnarray}
where $X_{\rm N_2}$ is the ratio of N$_2$ in the atmosphere, $\rho_0$ is the atmospheric density, $\rho_{\rm t}$ is the planetary \textcolor{black}{mean} density, and $R_{\rm t}$ is the planetary radius.  \textcolor{black}{In the derivation, we used the definition of the scale height $H=\frac{k_{\rm B}T}{\bar{m}g}$, where $k_{\rm B}$ is the Boltzmann constant, and $g$ is the gravitational acceleration.} The ratio of the N$_2$ mass to the planetary mass is proportional to ${\rho_0}/({R_{\rm t}^2\rho_{\rm t}^2})$. 
Equation \ref{eq:N2massPlanetmass} shows that the scaled N$_2$ mass on each planet is determined by $\rho_0/(R_{\rm t}^2 \rho_{\rm t}^2)$. \par

Here we analytically estimate the difference in the scaled N$_2$ mass between Earth and Mars. Our standard model showed that $\rho_0$ in the steady state is $\sim$4.4 times smaller on Mars than on Earth. The difference is likely to be caused by the smaller gravity of Mars. In contrast, the denominator ${R_{\rm t}^2\rho_{\rm t}^2}$ is $\sim$6.9 times smaller on Mars than on Earth. Consequently, the scaled N$_2$ mass $m_{\rm N_2}/M_{\rm t}$ on Mars is $\sim$1.6 times larger than that on Earth. This means that, even though the small gravity of Mars resulted in the smaller $\rho_0$, the scaled N$_2$ mass became larger because Mars has a larger ratio of surface area to planetary mass than Earth.\par

In other words, for a given $m_{\rm N_2}/M_{\rm t}$, a larger planet has a larger $\rho_0$ because of the smaller ratio of the surface area to the planetary mass. The larger $\rho_0$ allows more efficient removal of the atmosphere due to atmospheric erosion. As a result, the larger the planet we assumed, the smaller $m_{\rm N_2}/M_{\rm t}$ we obtained in the steady state. This is the reason why Mars obtained the largest amount of N$_2$ in our model.

\subsection{Analysis of the effect of element partitioning}\label{sec:negativefeedback}
Next, we discuss why the effect of element partitioning was limited in our numerical model. 
In the calculations, we found a relationship between the N$_2$ mass and the total atmospheric mass at the steady state.
\begin{figure}[htbp]
 \centering
 \includegraphics[width=7.5cm,clip]{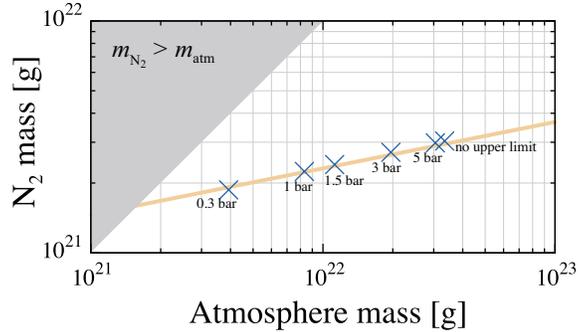}
 \caption{N$_2$ mass in the steady state $m_{\rm N_2, steady}$ as a function of total atmospheric mass at the steady state $m_{\rm atm, steady}$ for the Earth model varying the upper limit of CO$_2$ partial pressure as $P_{\rm CO_2}^{ \rm crit}$= 0.3 bar, 1 bar, 1.5 bar, 5 bar, and the case without upper limit of CO$_2$ partial pressure. The yellow line is the linear fitting line. The gray area represents an impossible situation ($m_{\rm N_2}>m_{\rm atm}$).}
 \label{fig:N2massMatmrelation} 
\end{figure}
Figure \ref{fig:N2massMatmrelation} shows the relationship of the N$_2$ mass and the atmospheric mass at the steady state in log-scale. The power  \textcolor{black}{law} index of the slope was $\sim$0.2, which means that the dependency on the atmospheric mass is, 
\begin{eqnarray}
m_{\rm N_2, steady} \propto m_{\rm atm, steady}^{0.2},
\label{eq:etamatm}
\end{eqnarray}
where $m_{\rm N_2, steady}$ and $m_{\rm atm, steady}$ are the N$_2$ mass and atmospheric mass at the steady state, respectively. 
The balance between the volatile supply and loss in the steady state should satisfy the following equation, 
\begin{eqnarray}
(1-\zeta)x_{\rm N_2}=\eta\cdot\frac{m_{\rm N_2, steady}}{m_{\rm atm, steady}}.
\label{eq:equilibrium}
\end{eqnarray}
The volatile abundance in an impactor $x_{\rm N_2}$ is constant and the impactor's escaping efficiency $\zeta$ is almost independent of $m_{\rm atm}$ in our model. The N$_2$ mass at the steady state $m_{\rm N_2, steady}$ on the right-hand side is proportional to $m_{\rm atm}^{0.2}$, and Equation \ref{eq:equilibrium} suggests that the atmospheric erosion efficiency $\eta$ is proportional to $m_{\rm atm}^{0.8}$. \par

Here, let us think about the behavior of the atmospheric evolution in our model. Early in the late accretion, the amount of atmosphere was still small and increased rapidly because the atmospheric supply was dominant. Later, the volatile supply and the loss were balanced. Since the volatile abundance in impactors was assumed to be constant and the impactor's escaping efficiency is almost independent of the atmospheric mass, the atmospheric erosion efficiency should have increased with the increase of atmospheric mass to reach the steady state.\par

\begin{figure}[htbp]
 \centering
 \includegraphics[width=7.5cm,clip]{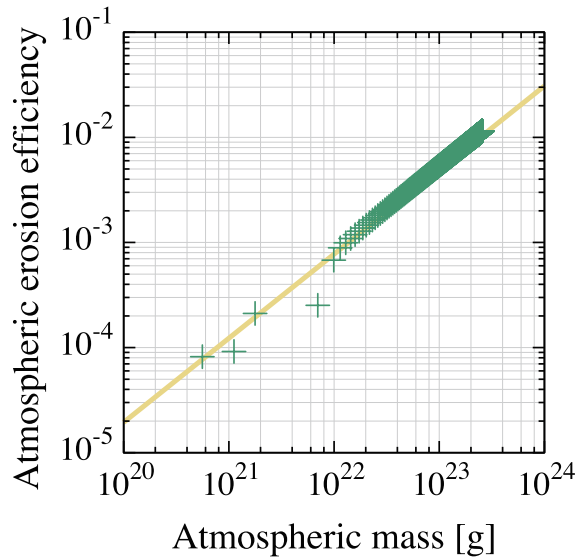}
 \caption{The dependency of the atmospheric erosion efficiency $\eta$ on the atmospheric mass $m_{\rm atm}$. The green plots correspond to the value of $\eta$ in the Earth model without upper limit of partial pressure, and the yellow line is the linear fitting line. }
 \label{fig:EtaMatmrelation} 
\end{figure}

Figure \ref{fig:EtaMatmrelation} shows the relationship of the $\eta$ and the atmospheric mass $m_{\rm atm}$ in log-scale. Indeed, the atmospheric efficiency increased with atmospheric mass, and the slope was 0.8 as expected. 
We note that, if each impact has removed part of the atmosphere in a geometrically identical figure, this slope should be 1. A smaller slope of 0.8 suggests that the larger amount of atmosphere reduced the escaping area. \par
Introducing the element partitioning in our model increased the N$_2$ abundance in the atmosphere and enhanced N$_2$ removal. However, the elemental partitioning also decreased the atmospheric mass, which resulted in a decrease in the efficiency of atmospheric erosion. These two competing effects led to the limited influence of element partitioning on N$_2$ mass as observed in our model.

\subsection{The survival of the primordial atmosphere on Venus}\label{sec:initialpressure}
We assumed a thin initial atmosphere with the total pressure $P=$ 0.1 bar in our model. 
In the steady state, where the partial pressure profiles show horizontal lines, the atmosphere is suggested to have almost been completely exchanged. Therefore, the contribution of the initial atmospheric pressure to the final pressure is found to be small and the volatiles provided in the late accretion should be dominant in the present-day atmosphere. However, the results also showed that the atmospheres on the terrestrial planets in the steady state do not reproduce the differences in noble gases and N abundances between Venus, Earth, and Mars.\par

\begin{figure}[htbp]
 \centering
 \includegraphics[width=7.5cm,clip]{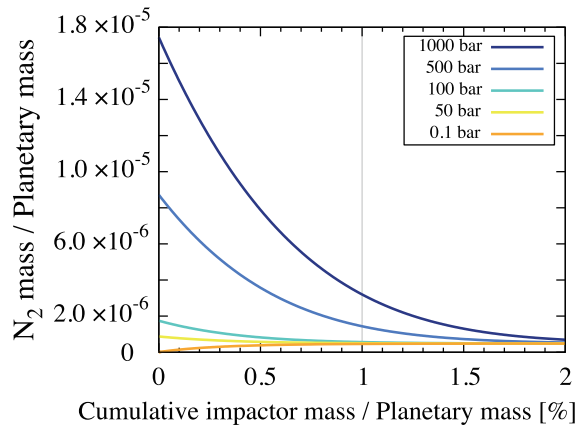}
 \caption{Dependences of final N$_2$ mass on the initial atmospheric pressure on Venus. The line color corresponds to the initial total pressure: 0.1, 50, 100, 500,   \textcolor{black}{and} 1000 bar. The impactor composition: $X_{\rm C}=0.1$, which corresponds to the enstatite chondrite-like impactor. Settings are same as Venus standard model. 1\% of the planetary mass (gray vertical line) corresponds to the end of the late accretion.}
 \label{fig:VenusInitialPressure_Echond} 
\end{figure}

One possible scenario to explain the difference between Venus and our model is the survival of the primordial atmosphere on Venus \cite[][]{pepin1991, gendaabe2005}. In order to investigate the condition for the primordial component to survive during the late accretion, we calculated the atmospheric evolution on Venus in the models where we started from the higher initial atmospheric pressure. 
Figure \ref{fig:VenusInitialPressure_Echond}  \textcolor{black}{shows} the N$_2$ mass in the cases with the initial pressure from 0.1 bar to 1000 bar. 
N$_2$ mass profiles for each initial condition were converging on the steady state in both figures. The N$_2$ masses approached the steady value regardless of the initial condition. 
However, we note that the N$_2$ mass has not completely reached the steady state at the end of the late accretion. In the case starting with a 500 bar atmosphere, the N$_2$ abundance at the end of the late accretion was more than twice as large as that of the case with 0.1 bar. This result suggests that the survival of the primordial atmosphere through the late accretion may have significant influence in the existence of the noble gases and N enrichment in the atmosphere of Venus. \par

\subsection{The effect of the existence an ocean}
\cite{gendaabe2005} demonstrated that the atmospheric loss during a giant impact was enhanced due to the presence of an ocean on early Earth. A ground motion by an impact induces complete vaporization of the ocean and the vapor pushes out the atmosphere. The CO$_2$-ice and H$_2$O-ice on Mars might experience the same process as an ocean.
While we neglected it in our model, the effect of ocean and ice vaporization on atmospheric erosion might have also been significant during the late accretion. 

\subsection{Atmospheric escape due to solar extreme UV and wind}

The difference in the abundances of noble gases and N between current Mars and our model might be caused by atmospheric escape due to solar extreme UV and wind. Noble gases (Ne, Ar, and Xe) and N in the Martian atmosphere have been known to be isotopically fractionated, which suggests the contribution of these atmospheric escape processes \cite[][]{jakosky2001, jakosky2017, kurokawa2018}. The atmospheric evolution models considering such isotopic fractionation suggested that Mars had a dense atmosphere whose atmospheric pressure was higher than 0.5 bar at 4 Ga \cite[][]{kurokawa2018}. This amount of atmosphere can be removed by ion pick-up sputtering induced by the solar wind \cite[][]{jakosky2017}.
 \textcolor{black}{
We note that hydrodynamic escape caused by solar extreme UV might have also influenced noble gases and N budgets on Earth and Venus, though their heavy-isotope enrichment is less remarkable than the case of Mars \cite[e.g.,][]{pepin1991,gillmann2009}.
}

 \textcolor{black}{
\subsection{Upper limit of the impactor size distribution}
In our model, the impactor size distribution is proportional to impactor diameter as $\mathrm{d}N(D)=\mathrm{d}D / D^3$. We assumed the maximum size of $D$ = 1000 km. In the assumed impactor-size distribution and total mass, the size of impactor that collides just once is $D\sim$900 km for Earth and Venus and $D\sim$500 km for Mars. Therefore, our statistical treatment is fair for Earth and Venus. }\par
 \textcolor{black}{In the case of Mars,  the lower maximum size of impactors might have slightly enhanced the atmospheric erosion on Mars as small impactors dominate the atmospheric erosion, whereas the supply of volatile exceeds the erosion for large impactors \cite[][]{schlichting2015}. We found that assuming the maximum size of $D$ = 500 km reduced the resulting N$_2$ mass on Mars by a factor of $\sim$2, which is insufficient to reproduce the gaps between three planets. Therefore, it does not influence our conclusions.}

 \textcolor{black}{
\subsection{Volcanic and impact-induced degassing}\label{sec:volcanic}
Volcanic and impact-induced degassing were not treated in our model, but they would hardly change the results.
The small dependence of the remaining N$_2$ mass on CO$_2$ and H$_2$O pressure limits suggests that the degassing of CO$_2$ and H$_2$O does not help to reconcile the gaps in noble gases and N.
The difference in the efficiency of N degassing might explain the gaps in N abundances between Earth and Venus \cite[][]{wordsworth2016}.
However, because the pattern of noble gas abundances in Earth's mantle is different from that in the atmosphere \cite[][]{marty2012}, the gaps in noble gases cannot be explained. We note that volcanic degassing of CO$_2$ was implicitly assumed for Earth as a part of carbon-silicate cycle.
}

\subsection{Various scenarios of planetary and atmospheric formation}
There are various planet formation scenarios such as the traditional model, Nice model, and Grand Tack model \cite[e.g.,][]{kokuboida1998, walsh2011}.\par
\cite{hansen2009} built a successful planetary formation model with all of the mass initially confined to a narrow annulus between 0.7 and 1.0 au, where Venus and Earth are orbiting. Mercury and Mars can form from material diffusing out of the annulus in their model. In this scenario, fewer objects would have hit Mars compared to Venus and Earth, suggesting a smaller amount of noble gases and N might have been obtained. The distributions of impact velocity might also differ from our model. The ambiguity of input parameters is still controversial.

By considering the ambiguity of input parameters and the dependencies on parameters showed in the subsections \ref{sec:TandPcrit} and \ref{sec:Cchond}, we calibrated the initial condition to reproduce the abundances of noble gases and N of the present-day atmospheres on the terrestrial planets.\par

\begin{figure}
 \centering
 \includegraphics[width=7.5cm,clip]{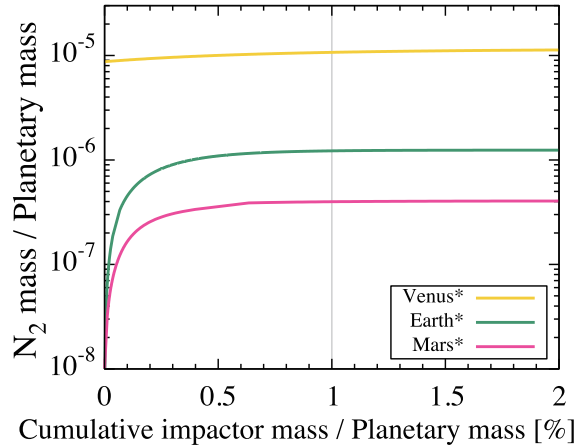}
 \caption{Same as Figure \ref{fig:Std3PlanetsN2mass}, but for the case that assumed special conditions on each planet: low surface temperature $T=2000$ K and volatile-rich impactor ($X_{\rm C}=1$) for Venus, a medium composition impactor ($X_{\rm C}=0.3$) for Earth, low CO$_2$ critical pressure limit $P^{\rm crit}_{\rm CO_2}=6$ mbar and volatile-poor impactor ($X_{\rm C}=0.1$) for Mars.}
 \label{fig:OnePossibleICofVEM}
\end{figure}

Figure \ref{fig:OnePossibleICofVEM} shows the N$_2$ mass evolutions of Venus, Earth, and Mars setting special initial conditions based on one possible scenario. We assumed different compositions of the late accretion impactors for each planet by setting the abundance of volatiles in an impactor to $X_{\rm C}=1$ on Venus, $X_{\rm C}=0.3$ on Earth, and $X_{\rm C}=0.1$ on Mars. Furthermore, several additional restrictions on initial conditions were assumed as follows: low surface temperature $T=2000$ K and high initial total atmospheric pressure of 500 bar on early Venus, and low upper limit of CO$_2$ partial pressure $P^{\rm crit}_{\rm CO_2}=6$ mbar on early Mars. \par 
In this special case, compared to Earth, Venus acquired roughly one order of magnitude more, and Mars was provided with $\sim$70\% less N$_2$. 
This result suggests that the differences in the abundances of noble gases and N in the atmospheres on Venus, Earth, and Mars can be reproduced by considering the special conditions.  
On the contrary, asteroids falling on Mars are thought to be richer in volatiles than those on Venus or Earth in the standard planet formation theory. Therefore, some additional mechanisms are needed to realize such special conditions, which provides different composition impactors to each terrestrial planet.\par

\textcolor{black}{We note that the special conditions discussed above were required chiefly because of Mars. The gaps in the abundances of N and noble gases between Venus and Earth can be at least partially reproduced by a lower surface temperature on Venus and a smaller upper limit of CO$_2$ partial pressure on Earth (subsection \ref{sec:TandPcrit}). However, the low abundances on Mars cannot be reproduced without the difference in the impactor composition in our model (Figure \ref{fig:OnePossibleICofVEM}). The difficulty to reproduce Martian atmosphere was mainly caused by the  large ratio of the surface area to the planetary mass (subsection  \ref{sec:smallmars}).}

\section{Conclusions}\label{sec:conclusion}
We investigated the effect of element partitioning at the surface of early terrestrial planets on atmospheric evolution. We modeled the impact degassing and atmospheric erosion during the late accretion. In our model, a state of runaway greenhouse on Venus, carbon-silicate cycle and existence of oceans on Earth, and CO$_2$-ice and H$_2$O-ice formation on Mars were assumed by setting upper limits to the partial pressures of CO$_2$ and H$_2$O on Earth and Mars. 
The amount of noble gases and N obtained at the steady state decreases by $\sim$40\% and $\sim$15\% for Earth and Mars respectively due to the effect of element partitioning. The effect alone was found to be insufficient explanation for the distinct differences -- roughly two orders of magnitude -- in the abundance of noble gases and N between the atmospheres of Venus, Earth, and Mars. 
The element partitioning enhanced N$_2$ removal and it also decreased the atmospheric mass. The atmospheric erosion efficiency depends on the atmospheric mass and so the influence of elemental partitioning on N$_2$ mass was limited.\par
As a result, the amount of N$_2$ obtained by Mars was about twice as large as that of Venus and Earth. We found that the N$_2$ mass scaled by planetary mass is determined by $\rho_0/(R_{\rm t}^2 \rho_{\rm t}^2)$. Therefore Mars obtained a small amount of N$_2$ due to the larger ratio of surface area to planetary mass. 
The atmospheric evolution also depends on input parameters. 
The high temperature and low upper limits of the partial pressures case resulted in less remaining N$_2$ remained. 
However, it is difficult to reproduce the distinct gaps between the abundances of noble gases and N in the three planets' present-day atmospheres even if we considered a wide parameter space.
This suggests that the survival of the primordial atmosphere through the late accretion on Venus, and the atmospheric escape by solar extreme UV and wind on Mars should partially account for the present-day atmospheres.

\appendix

\section{Size distribution dependence}
Since the origin of the late accretion impactors is unknown, the size distribution of the impactors was also treated as a parameter. We assumed the projectile size distribution $\mathrm{d}N/\mathrm{d}D\propto D^{-p}$, where $\mathrm{d}N$ is the number of objects of diameter $D$ within a bin $\mathrm{d}D$, and $p$ is the power law index. According to the model by \cite{bottke2005}, the size frequency distribution of the asteroid belt is roughly proportional to $D^{-3}$($p=3$) and we used this distribution in our model.\par

\begin{figure}
 \centering
 \includegraphics[width=7.5cm,clip]{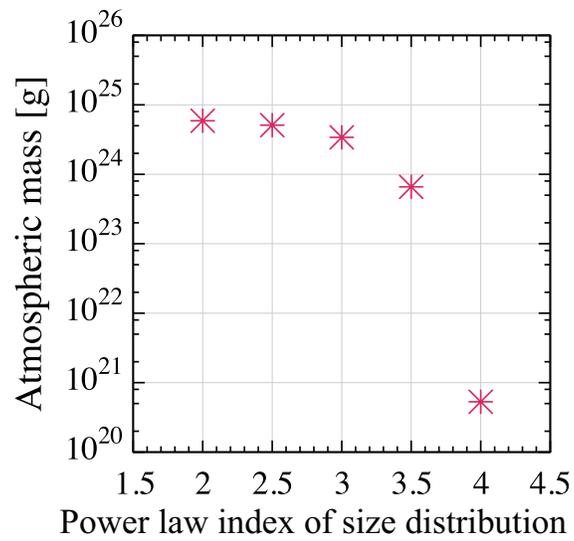}
 \caption{Dependence of final atmospheric mass on the impactor size distribution. The power law index of impact size distribution $p$ is defined as $\mathrm{d}N(D)/\mathrm{d}D\propto D^{-p}$ and this figure plotted the atmospheric mass at the point of 1\% of planetary mass impact for $p=2.0,\ 2.5,\ 3.0,\ $3.5, and 4.0. The value $p=3.0$ corresponds to the size distribution of the present-day main belt asteroids \cite[][]{bottke2005}.}
 \label{fig:SizeDpowerindex}
\end{figure}

We varied the value of the power law index $p=2.0, 2.5, 3.0, 3.5,$  \textcolor{black}{and} 4.0. Low $p\ (<3)$ values correspond to shallow size distributions where most of the mass of the impact population is in the large projectiles, whereas high $p\ (>3)$ values correspond to a steep slope where the small projectiles dominate the total impactor mass. Figure \ref{fig:SizeDpowerindex} shows the resulting atmospheric masses of these simulations for Earth. For other input parameters, we used the values of the Earth model. 
The steep size distribution impactor with high $p\ (>3)$, containing a large amount of small projectiles, strongly eroded the vapor plume and left a thin atmosphere behind. Considering the impactor size distribution with $p=4.0$, the resulting atmospheric mass at the point of 1\% of planetary mass impact -- which corresponds to the end of the late accretion -- was a much smaller amount than that for $p=3.0$ by several orders of magnitude. This is due to the dependence of atmospheric erosion on the impactor size. We found that an impactor of several kilometers in diameter is most efficient for atmospheric erosion due to maximizing the value of atmospheric erosion efficiency $\eta$.

\section*{Acknoledgement}
We thank Satoshi Okuzumi, Shigeru Ida, Masahiro Ikoma, Hidekazu Tanaka, and Takanori Sasaki for valuable discussions.
We acknowledge the financial support of MEXT KAKENHI grant (JP17H06457  \textcolor{black}{and 15J09448}).

\section*{References}

\bibliography{mybibfile}

\end{document}